\documentclass[]{aiaa-tc}

\usepackage[toc]{glossaries}
 \usepackage{varioref}
 \usepackage{wrapfig}
 \usepackage{threeparttable}
 \usepackage{dcolumn}
  \newcolumntype{d}{D{.}{.}{-1}}
 \usepackage{nomencl}
  \makeglossary
 \usepackage{subfigure}
 \usepackage{subfigmat}
 \usepackage{fancyvrb}
  \fvset{fontsize=\footnotesize,xleftmargin=2em}
 \usepackage{lettrine}
 \usepackage[dvips]{dropping}
 \usepackage[colorlinks]{hyperref}
\usepackage{amsmath}
\usepackage{graphicx}
\usepackage{epstopdf, epsfig}
\usepackage{color,macros}
\graphicspath{{figures/}}
\usepackage{multirow}
\usepackage{tabularx}

\newcommand{\eg}{\emph{e.g.},~}

 \title{Separation delay via hydro-acoustic control of a \nacafft airfoil in pre-stalled conditions}

 \author{
Julien Bodart$^{\mathrm{a},}$\thanks{Assistant Professor, IASE-Supaero, Toulouse France},
 Carlo Scalo$^{\mathrm{b},}$\thanks{Assistant Professor, School of Mechanical Engineering, Purdue University, USA}, 
 Grigory Shelekhov$^{\mathrm{a}}$\thanks{Graduate Research Assistant, IASE-Supaero, Toulouse France}$\,$ and Laurent Joly$^{\mathrm{a}}$\thanks{Department Head, IASE-Supaero, Toulouse France} \\
  {\normalsize\itshape
   $^\mathrm{a}$~DAEP, ISAE-Supaero, BP 54032, 31055 Toulouse Cedex 04, France} \\
  {\normalsize\itshape
   $^\mathrm{b}$~School of Mechanical Engineering, Purdue University, West Lafayette, IN 47907-2088, U.S.A.}\\
  }
  
 \AIAApapernumber{2017-NUMBER}
 \AIAAconference{SCITECH, Jan. 2017, and Grapevine, Texas}
 \AIAAcopyright{\AIAAcopyrightD{2017}}

\begin{document}

\maketitle

\begin{abstract}

We have performed large-eddy simulations of turbulent separation control via impedance boundary conditions (IBCs) on a \nacafft airfoil in near-stalled conditions. The uncontrolled baseline flow is obtained for freestream Mach numbers of $M_\infty=0.3$, chord-Reynolds numbers $Re_c = 1.5\times10^6$ and angle of attack, $\alpha=14^{\circ{}}$ where separation is obtained on the suction at approximately 85\% of the chord.  The adopted flow conditions are inspired by the experimental setup of Coles \& Wadcock (1979).

Flow control is applied via imposition of complex IBCs using the time-domain implementation developed by Scalo, Bodart, and Lele, \emph{Phys. Fluids} (2015). The adopted IBCs model an array of sub-surface-mounted Helmholtz cavities with tunable resonant frequency, $f_{res}$, covered by a porous sheet with permeability inversely proportional to the impedance resistance $R$. Separation is delayed due to the enhanced mixing associated with convectively amplified spanwise-oriented Kelvin-Helmholtz (KH) rollers, generated via hydro-acoustic instabilities. The latter are the result of the interaction of the wall-normal transpiration through the impedance panel (sustained by acoustic resonance, modeled by the IBCs) and the overlying mean background shear. The KH rollers initial size, before convective amplification, is determined by the period of the passively controlled oscillations in the transpiration velocity. It hence scales as $l_{KH,0} \simeq U_\infty/f_{res}$, where $U_\infty$ is the free-stream velocity.

Panel resonant reduced frequencies, based on the panel resonant frequency $f_{res}$, length of the separated region, $\l_\textrm{sep}$ and free-stream velocity $U_\infty$, have been varied in the range of $F^+_{res}=0.14-1.34$. For each of these frequencies, impedance resistances in the range $R=0.1 -- 0.5$ were tested. The result is an alteration of the coupled instability between the separating shear layer and the vortex shedding in the wake (already present in the uncontrolled baseline flow) yielding unique wake topologies associated with different intensities for the passively generated KH vortical structures. Specifically, enhancements up to +13\% in the lift coefficients have been obtained for $F^+_{res} \simeq 0.3$. Results show that tuning of the resonant cavities below the natural shedding frequency is required to generate KH rollers structures with a sufficiently large entrainment diameter to encompass the full extent of the separated region, thereby enhancing mixing and promoting reattachment.  Overall, the results presented in this work show that the adoption of hydro-acoustically tuned resonant panels is a promising passive control technique for boundary layer separation control.
\end{abstract}

\printglossary
 

\section{Introduction}

\lettrine[nindent=0pt]{T}{he} present work explores the possibility of using wall-mounted resonating porous panels, modeled via impedance boundary conditions (IBCs), as a means to passively control the overlying turbulent boundary layer. The fundamental working principle of such control involves acoustic resonance in the panel, being triggered by the energy containing turbulent eddies, generating a wall-normal transpiration pattern at the wall. The latter then generates Kelvin-Helmholtz (KH) rollers due to the presence of the background mean shear. We refer to this phenomenon as \emph{hydro-acoustic instability}, since it involves the coupling between acoustic resonance and classic hydrodynamic instability. This passive flow control principles has been exploited in the current work to passively control turbulent separation delay. To the author's knowledge, it is the first time that such topic is tackled with the use of high-fidelity numerical simulation tool such as large-eddy simulation coupled with time-domain impedance boundary conditions (TDIBCs) as developed by Scalo, Bodart and Lele, \emph{Phys. Fluids} (2015).

Previous related work on the topic mainly focuses on \emph{active} control. Nishioka et al.\cite{nishioka1990control} has reduced the extent of the separated region over a stalled flat plate at a high angle of attack via imposition of sound waves at the leading edge. Huang et al.\cite{huang1988effect} investigated the effects of active acoustic forcing on trailing edge separation and near wake development on a symmetric airfoil at zero angle of attack at low chord Reynolds numbers $Re_c = 3.5 \times 10^4$, achieving significant separation delay by introducing a slot near the separation point.
Dandois et al. \cite{dandois2007numerical} have investigated separation control of a turbulent compressible boundary layer evolving over a rounded back-facing step where transpiration was imposed upstream the separated shear layer through the neck of a Helmholtz-like cavity actively controlled via imposed oscillations of the cavities' bottom panel.

To the authors knowledge, there are only a handful of published experimental efforts on passively controlled flow separation over an airfoil. Yang \& Spedding\cite{yang2013passive} placed resonating cavities along the suction side of a stalled E387 airfoil in a low-Reynolds number and low-Mach number flow settings successfully controlling laminar separation.
Urzynicok \cite{urzynicok2003separation} investigated the interactions between flow and acoustic resonance as possible means to passive separation control. An airfoil FX 61-184 was placed at a high angle of attack and laminar separation control was studied. The passive control was also designed for the turbulent boundary layer developing in a diffuser.

The aforementioned experiments have inspired the current effort, which targets high Reynolds numbers flows at moderate Mach numbers. The baseline uncontrolled separated flow used in the present numerical study is the compressible counterpart of the pre-stalled flow measurements by Coles \& Wadcock \cite{coles1979flying}. The choice for these experiments comes from their wide-spread use within the CFD community to validate RANS models against boundary layer separation at high Reynolds on airfoils\cite{NasaTurb}. In their experiment, trailing-edge separation occurred for $\alpha \ge 14^{\circ}$ and it has been well documented with a flying hot wire measurements, later completed by several studies within the same wind tunnel as well as other facilities \cite{hastings1984studies}.


In the following we first introduce the setup under investigation (section \ref{sec:prob_formulation}) and the computational tools adopted to perform high-fidelity simulations (section \ref{sec:comp_setup}), where the characteristics-based implementation of IBCs is explained, as well as the underlying numerical scheme. The uncontrolled pre-stalled flow is then analyzed (section \ref{sec:pre_stalled_flow}) with particular focus on a time scale analysis of turbulence and the separated flow, as well as grid convergence study and sensitivity of the results to the spanwise-extend of the domain. The design of the control strategy based on resonating panels is described in section \ref{sec:control_strategy} and results from large-eddy simulations are discussed in section \ref{sec:results}.

\section{Problem Formulation} \label{sec:prob_formulation}

\begin{figure}
  \centering
    \includegraphics[width=\textwidth]{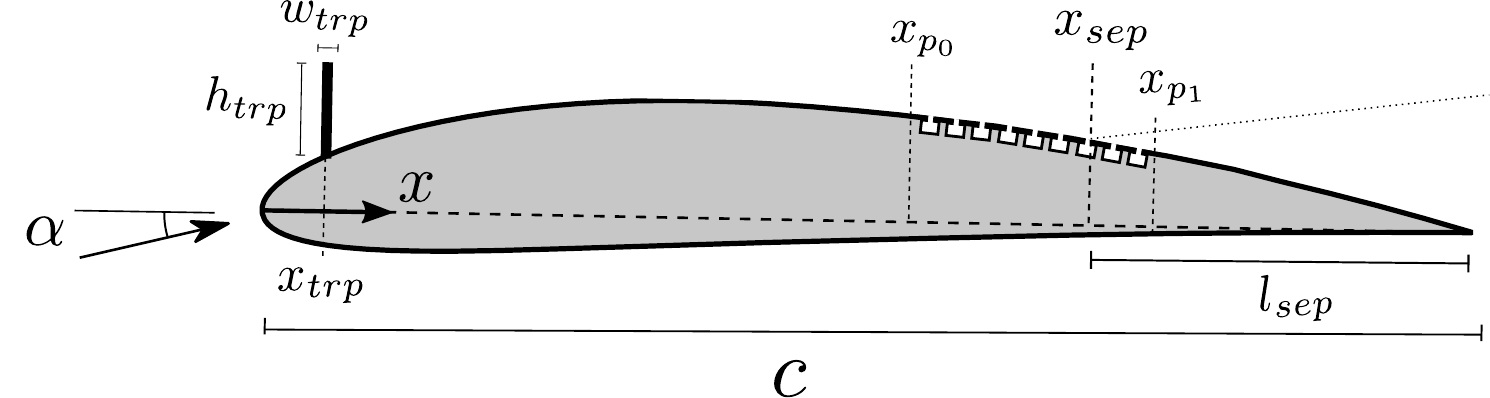}
\begin{center}
\centering
\begin{tabularx}{0.8\textwidth}{m{1.3cm}m{1.2cm}m{1.38cm}m{1.38cm}m{1.1cm}m{1.1cm}m{1.1cm}m{1.1cm}}
 $\alpha$ &  $x_{trp}/c$ & $h_{trp}/c$ & $w_{trp}/c$ & $x_{p0}/c$ &  $x_{p1}/c$  & $x_{sep}/c$ & $l_{sep}/c$   \\ 
\hline
 14.5$^{\circ}$ & 0.093 & $2\times10^{-3}$ & $5\times10^{-5}$ & 0.55  & 0.75  & 0.77 & 0.23
\end{tabularx}
\end{center}
    \caption{Computational setup of the controlled case showing the NACA-4412 airfoil geometry at angle of attack, $\alpha$, free-stream velocity, $U_\infty$, and chord, $c$. The approximate location of the separated shear-layer of the uncontrolled case is shown with a dotted line. IBCs are applied between $x_{p0} < x < x_{p1}$ on the suction side, partly overlapped with the separated region, $x > x_{sep}$ (heuristically determined, suboptimal location). IBCs are symbolically represented by an array of Helmholtz resonating cavities. The boundary layer on the suction side is tripped at location $x=x_{trp}$, as shown in the inset figure.}
 \label{fig:naca4412}
\end{figure}

The computational setup consists of a NACA-4412 airfoil equipped with an impedance panel (heuristically) located upstream of the separated region. In preliminary trials several locations of the impedance panel were tested, targeting the boundary layer on the suction side at different stages of development, from 
strong favorable pressure gradient (near the trip), adverse pressure gradient upstream of the separation zone and fully immersed in the separated region, $x>x_{sep}$ (\fig{fig:naca4412}). When a sufficiently strong background mean shear is present, the flow response is manifest in the form of a hydrodynamic Kelvin-Helmholtz (KH) instability triggered by wall-normal acoustic resonance modeled by the imposed IBCs. We discuss a subset of several simulations (Fig.~\ref{fig:naca4412}) for which the actuation is located right before the separated region where the generated KH rollers periodically reenergize the otherwise separated boundary layer, with a net gain in the lift coefficient, experiencing vortex-induced oscillations. The investigation hereafter focuses on analyzing the sensitivity of the flow response to various IBC parameters by fixing the impedance panel to the thus-found sub-optimal location.

\section{Computational Setup} \label{sec:comp_setup}

\subsection{Governing Equations}

The governing equations for mass, momentum, and total energy are solved in conservative form,
\begin{subequations}
	\label{eq:navierstokes}
	\begin{align}
		\frac{\partial}{\partial t} \left(\rho\right) &+ \frac{\partial}{\partial x_j} \left(\rho u_j \right)  = 0
		\label{subeq:ns1}
		\\
		\frac{\partial}{\partial t} \left(\rho u_i\right) &+ \frac{\partial}{\partial x_j} \left(\rho u_i u_j\right)  =  -\frac{\partial}{\partial x_i} p  +
		\frac{\partial}{\partial x_j} \tau_{ij}
		\label{subeq:ns2}
		\\
		\frac{\partial}{\partial t} \left(\rho \, E\right) &+ \frac{\partial}{\partial x_j} \left[ u_j \left(\rho \, E + p \right) \right] =
		\frac{\partial}{\partial x_j } \left(u_i \tau_{ij} - q_j\right)
		\label{subeq:ns3}
	\end{align}
\end{subequations}
where $x_1$, $x_2$, and $x_3$ (equivalently, $x$, $y$, and $z$) are the chord-wise, chord-normal and span-wise coordinates and $u_i$ are the velocity components in each of those directions, and $p$, $\rho$, and $E$ are respectively pressure, density, and total energy per unit mass. The gas is assumed to be ideal, with equation of state $p= \rho \,R_{gas}\, T$ and a constant ratio of specific heats, $\gamma$. The gas constant is calculated as $R_{gas} = p_{\textrm{ref}} \left(T_{\textrm{ref}}\,\rho_{\textrm{ref}}\right)^{-1}$, where $\rho_{\textrm{ref}}$, $p_{\textrm{ref}}$, and $T_{\textrm{ref}}$ are respectively the reference thermodynamic density, pressure, and temperature. Changing the reference thermodynamic state is needed when one or more dimensionless groups governing the flow physics need to be scaled to match a desired value as discussed below.

The viscous and conductive heat fluxes are:
\begin{subequations}
	\label{eq:heatfluxes}
	\begin{eqnarray}
		\tau_{ij} &=& 2 \mu \left[S_{ij} - \frac{1}{3} \frac{\partial u_k}{\partial x_k} \delta_{ij} \right]\\
		\label{subeq:hf1}
		q_j &=& -\frac{\mu\,C_p}{Pr} \frac{\partial}{\partial x_j} T
		\label{subeq:hf2}
	\end{eqnarray}
\end{subequations}
where $S_{ij}$ is the strain-rate tensor, given by $S_{ij}=(1/2) \left(\partial u_j/\partial x_i + \partial u_i /\partial x_j \right)$; $Pr$ is the Prandtl number; $C_p=\gamma\,R_\textrm{gas}/(\gamma-1)$ is the specific heat a constant pressure; and $\mu$ is the dynamic viscosity, given by $\mu = \mu_{\textrm{ref}}\left(T/T_\textrm{ref}\right)^n$, where $n$ is the viscosity power-law exponent and $\mu_{\textrm{ref}}$ is the reference viscosity. 
The baseline, uncontrolled simulations has been carried out with free-stream Mach number $M_\infty=0.3$ and chord Reynolds number $Re_c = 1.5\,\times\,10^6$. The choice of a non-zero free-stream Mach number, $M_\infty=0.3$ (different from the experimental value of $M_\infty=0.1$) was dictated by the need to contain the computational cost per simulation, allowing to explore a broader range of IBC settings. 
Only for the sake of direct comparison with the experimentally measured statistics, and hence validation of the computational model, the Mach number was lowered to $M_\infty=0.1$ while keeping the chord Reynolds number and the free-stream velocity at $Re_c = 1.5\,\times\,10^6$.

\subsection{Boundary conditions \label{sec:boundary_conditions}}

No-slip and adiabatic boundary conditions are used everywhere in the model. The Eulerian portion of the flux is on the control surface is modelled by imposing impedance boundary conditions $Z(\omega)$, is defined such as
\begin{equation} \label{eq:impedance_definition}
\hat{p}(\omega) = \rho_0\,a_0\,Z(\omega)\,\hat{v}_n(\omega)  \,
\end{equation}
between the Fourier transforms of pressure, $\hat{p}$, and the wall-normal velocity, $\hat{v}_n$ (positive if oriented away from the fluid side), where $\rho_0$ and $a_0$ are the base density and speed of sound. The use of IBC done this way significantly alleiviates the grid resolution requirement, as the pores and Helmholtz cavities do not need to be resolved, while allowing to retain high-fidelity on the flow side and on the modeling of the acoustic response at the boundary, simultaneously. Discussion on modeling issues like this one can be found in \cite{leschziner2011simulation}. In the present computations the impedance boundary conditions \eqref{eq:impedance_definition} have been imposed following \cite{ScaloBL_PoF_2015}'s implementation which relies on time-domain impedance formulation by \cite{FungJ_AIAAJ_2001,FungJ_IntJCFD_2004}.
In practice, impedance boundary conditions are not implemented as a direct relationship between pressure and velocity \eqref{eq:impedance_definition} but rather in the wave-space
\begin{equation} 
v_n^{\pm}(n,t) = \frac{p'(n,t)}{\rho_0\,a_0} \pm v_n'(n,t)
\end{equation}
in terms of outgoing $v_n^{+}(n,t)$ and ingoing $v_n^{-}(n,t)$ waves.

The impedance boundary conditions are implemented via the complex wall softness coefficient, $\widehat{\widetilde{W}}$,
defined as the complex reflection coefficient, $\widehat{W}_\omega(\omega)$, via
\begin{align} \label{eqn:reflection_coefficient}
\widehat{W}_\omega(\omega) \equiv \frac{Z_0-Z\left(\omega\right)}{Z_0+Z\left(\omega\right)} = \widehat{\widetilde{W}}_\omega(\omega) - 1\; ,
\end{align}
where $Z_0 = \rho_0\,a_0$ is the base impedance. Hard-wall (purely reflective) conditions are imposed by setting $\widehat{\widetilde{W}}=0$, corresponding to the limit of infinite impedance magnitude $|Z|\rightarrow\infty$.

\begin{figure}
\centering
\includegraphics[width=0.9\textwidth]{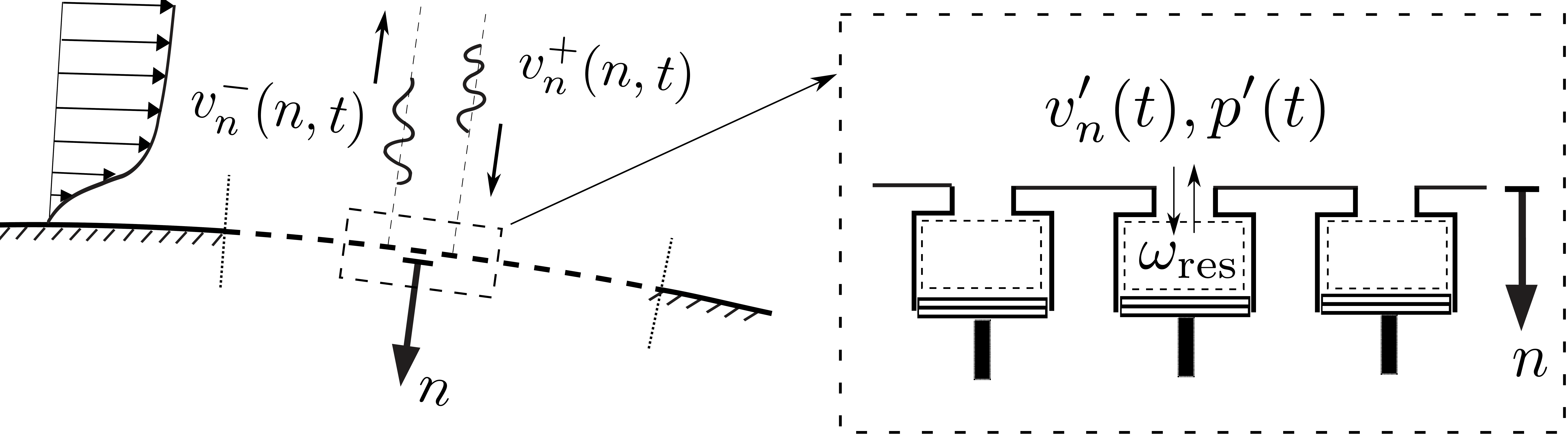}
\caption{Illustration of locally one-dimensional implementation of impedance boundary conditions, with impedance cast in form of permeability (a). Conceptual design of hydro-acoustically tuned porous surface where Helmholtz cavities sizes can be conceptually tuned with micro-actuators.}
\label{fig:IBC_sketch}
\end{figure}

A specific form of impedance boundary conditions, a single-oscillator Helmholtz resonator given by the expression \eqref{eq:single_oscillator_IBC}, will be adopted in the present calculation. Its physical meaning, illustrated in figure \ref{fig:IBC_sketch}, will be analyzed in section \ref{sec:control_strategy}, where the specifics of the resonance-based control strategy will be described.

\subsection{Numerics}

The governing equations are solved using \cx, a control-volume-based, finite-volume solver for the fully compressible Navier--Stokes equations on unstructured grids. \cx~employs a variable-stage Runge-Kutta time discretization and a grid-adaptive reconstruction strategy, blending a high-order polynomial interpolation with low-order upwind fluxes \cite{HamMIM_2007_bookchpt}. The code is parallelised using the Message Passing Interface (MPI) protocol and is highly scalable on a large number of processors \cite{BermejoBLB_IEEE_2014}. It has already been applied to a large variety of problems, including shock turbulence boundary layer interaction and high-Reynolds number flows \cite{BodartLM_AIAA_2013,bermejo2014confinement,larsson2015large}.

\section{Uncontrolled Pre-Stalled Flow} \label{sec:pre_stalled_flow}

To investigate the effects of the impedance model on lifting surfaces, and its potential for flow control, we first establish a baseline case using the experimental database of \cite{coles1979flying} who
characterized a \nacafft~airfoil in pre-stalled conditions ($\alpha=13.87^{\circ}$) and with a chord-based Reynolds number $Re_c=1.52~10^6$. This case has been widely used within the CFD community to validate RANS models against boundary layer separation at high Reynolds on airfoils (\eg \cite{NasaTurb}). In particular, trailing-edge separation occurring at $\alpha\approx14^{\circ}$ has been well documented using flying hot wires in the work of \cite{coles1979flying}, later completed by several studies within the same wind tunnel \cite{wadcock1987} as well as other facilities \cite{hastings1984studies}. Regarding the impedance boundary conditions interaction with a turbulent flow, this test case is attractive as it presents adverse pressure gradient, boundary-layer separation, and is relevant to trailing edge stall. Interestingly, this database has also been used for control purposes using lift enhancement devices, namely Gurney flaps and vortex generators \cite{storms1994lift}.

\begin{table}
\centering
\begin{tabular}{ccccccccc}
Name & $L_z/c$ & $N_z$ & $N_{y}^\textrm{BL}$ & $\tau_{expn}$ & $N_{x}^\textrm{SS}$ & $N_{tot}$ & $\Delta y/c$ (first cell)  \\
\colrule
 A05 & 0.05 & 100 & 100 & 1.02 & 972 & $34$M  & $1.5 \cdot 10^{-4}$\\
 A10 & 0.1  & 200 & 100 & 1.02 & 972 & $68$M & $1.5 \cdot 10^{-4}$  \\
 B05 & 0.05 & 100 & 100 & 1.03 & 972 & $34$M & $7.5 \cdot 10^{-5}$\\
 C05 & 0.05 & 200 & 100 & 1.03 & 1042 & $66$M & $7.5 \cdot 10^{-5}$\\
\botrule
  \end{tabular}
\caption{Selected grids for the grid sensitivity analysis and spanwise dependency of the baseline simulation
of the flow around a \nacafft~airfoil, at $Re_c=1.52~10^6$. $N_{y}^\textrm{BL}$ is the number of grid point across the boundary layer, $\tau_{expn}$ the associated cell size expansion, $N_{x}^\textrm{SS}$ and $N_z$ the number of grid points in the streamwise direction on the suction side and in the spanwise direction, respectively. }
 \label{table:grids}
\end{table}

\subsection{Numerical setup}

Simulations are first carried out at a Mach number $M_{\infty}=0.1$ (hereafter referred to as the \textit{nominal} case) and angle of attack $\alpha=13.87^{\circ}$. 
We first aim at establishing an reference case which matches the experimental setup, and provides an accurate description of the flow field, with a particular focus recirculation region at the trailing edge. This validation step is essential to establish a baseline case for our numerical experiment, which would have to both capture the main flow features while
requiring affordable computational power.
For all the considered cases, the computational grid is a bi-dimensional C-type grid extruded in the spanwise direction.
Because \cx~allows for hybrid meshing, a structured mesh was used in the region of interest, \ie~the boundary layer and a significant part of the wake region ($L_{wake}=1.5c$ starting from the trailing edge).
Outside of these regions, a unstructured quad grid was used to save 
computational resources.
To evaluate the grid sensitivity to the LES solution, three different grids are considered before spanwise extrusion. The number of cells $N_x$ in the streamwise direction is kept constant, while both the grid points density across the boundary layer in the wall normal direction and the size of the first cell has been varied compared to the cheapest grid resulting in grids A,B and C described in the \tab{table:grids}.

All the grids have been extruded in the $z$ direction with the same span extent of 5 percent and $N_z$ grid points, except for A10, which has a doubled span extent, and is used to validate the $z$-periodicity of the flow field. While not shown here, no significant difference are observed in the $A10$ case, regarding the size of the recirculation bubble, the profiles of the velocity field and Reynolds stresses. 


The computational domain is bounded by boundary conditions located $20c$ away from the airfoil, where free-stream boundary conditions are imposed, without circulation correction (see \eg \cite{thomas1986far}). No-slip wall boundary condition are applied on the airfoil.
To trigger transition from laminar to turbulent state in the boundary layer, narrow strips of tape are located on both the pressure and the suction side of the airfoil ($h^{trip}\approx 1.7~ 10^{-4}c$) \cite{wadcock1978flying}. In the computation, no trip was added on the suction side, as transition naturally develops similarly to the experiment at $x=0.025c$, while a numerical trip consisting of sponge-like condition (forcing the flow towards rest) was added at location $x=0.103c$ ($h^{trip}=0.002c$) on the pressure side, as in the experiments. However, due to the marginal grid resolution and reduced viscosity effects, very limited differences are observed on the boundary layer development on the pressure side when compared to untripped simulations.

For each simulation, preliminary runs are designed using several intermediate level of refinement (not mentioned here) before computing the flow field on the various grid presented in \tab{table:grids}.
For each intermediate grid, the solution is advanced up to an estimated statistically steady state, following a minimum of $T_{transient}=2c/U_{\infty}$ of physical time.
Statistics are then sampled during a minimum of $T_{stats}=4c/U_{\infty}$. 

\subsection{Comparison with experimental data}

\begin{figure}
  \centering
  \includegraphics[width=0.47\textwidth]{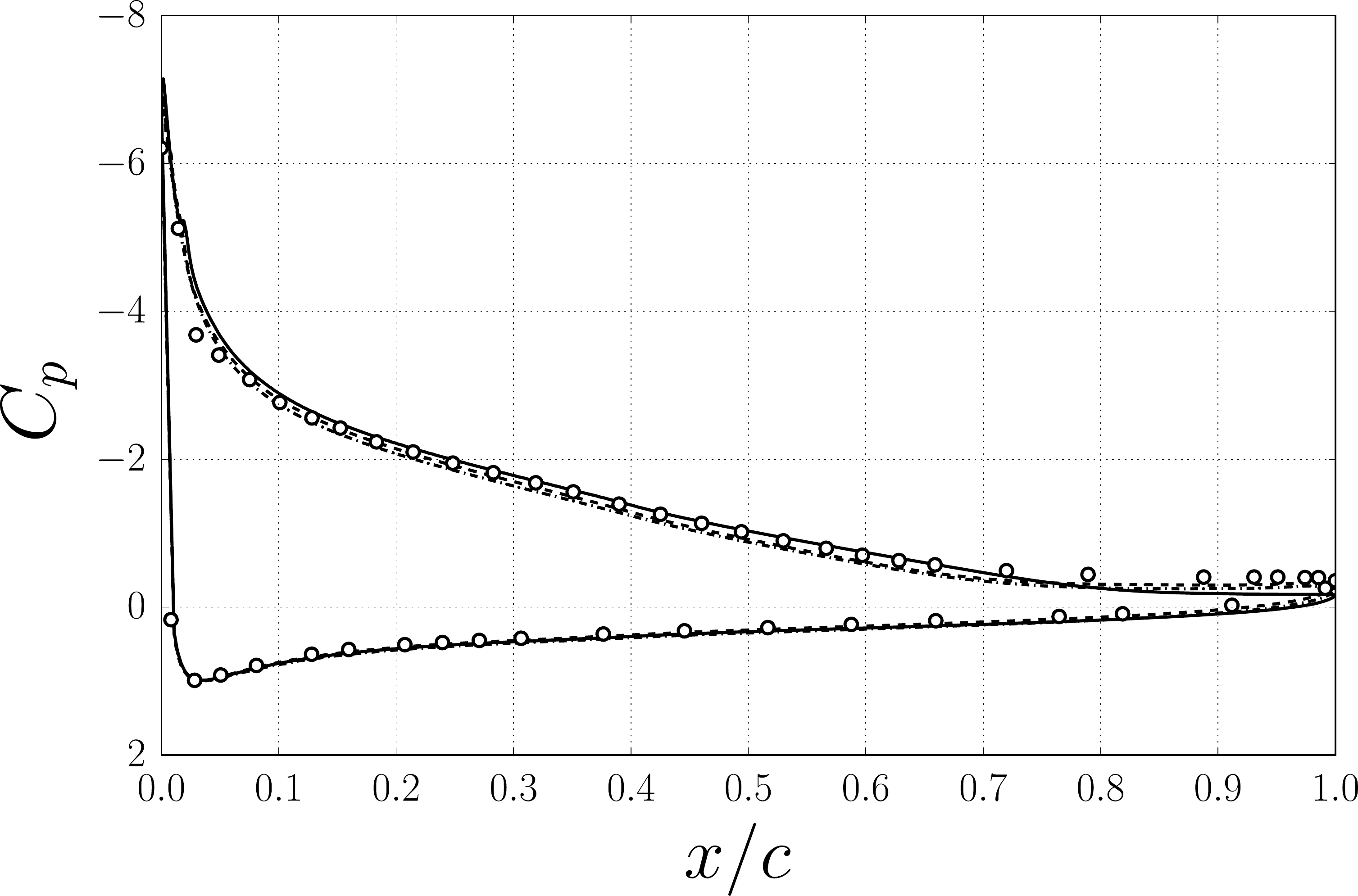}
  \includegraphics[width=0.47\textwidth]{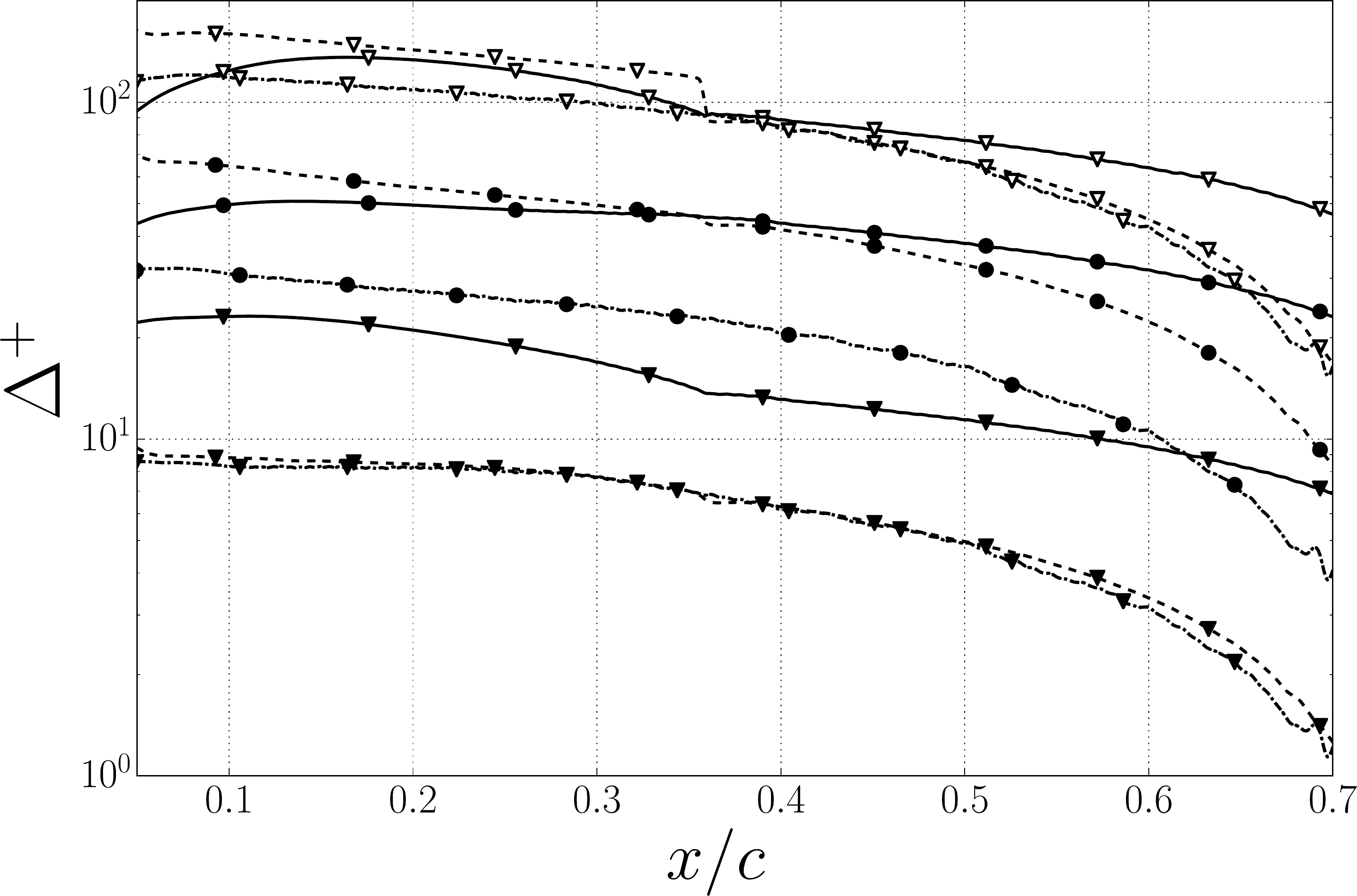} 
  \caption{\label{fig:Cp_validation} (a) Pressure coefficient $C_p$ for the 
  nominal case with grids listed in the table \ref{table:grids}. (\circle) experiments by \cite{coles1979flying}, \legendbaseline. (b) Evolution of the first off-wall cell size in wall units along the suction side. Legend: (\trianopd) $\Delta x^+$, (\triansold) $\Delta y^+$, and  (\solidcircle) $\Delta z^+$. 
   }
\end{figure}


The results are compared with available experimental results of \cite{wadcock1978flying}.
A comparison of the pressure coefficient $C_p$ is given on the \fig{fig:Cp_validation}. 
A very good agreement is obtained with experimental results for all the considered grids, especially regarding the intensity
of the adverse pressure gradient at the suction side, which demonstrates the negligible effect of the ``free air'' setup
as compared to the wall bounded wind tunnel environment. However, the pressure coefficient is slightly overestimated with the finest grid. This feature has also been found in ``free air'' RANS simulations \cite{NasaTurb} and plays a significant role in the prediction of the correct separation point $x_\textrm{sep}$.

The flattening towards the trailing edge is the footprint of the boundary layer separation. A significant difference is thus observed between runs A05 and B05, in which boundary layer separation is obtained for $x=0.87$ and $x=0.75$, respectively.

\begin{figure}
  
  \centering
   \includegraphics[width=\textwidth]{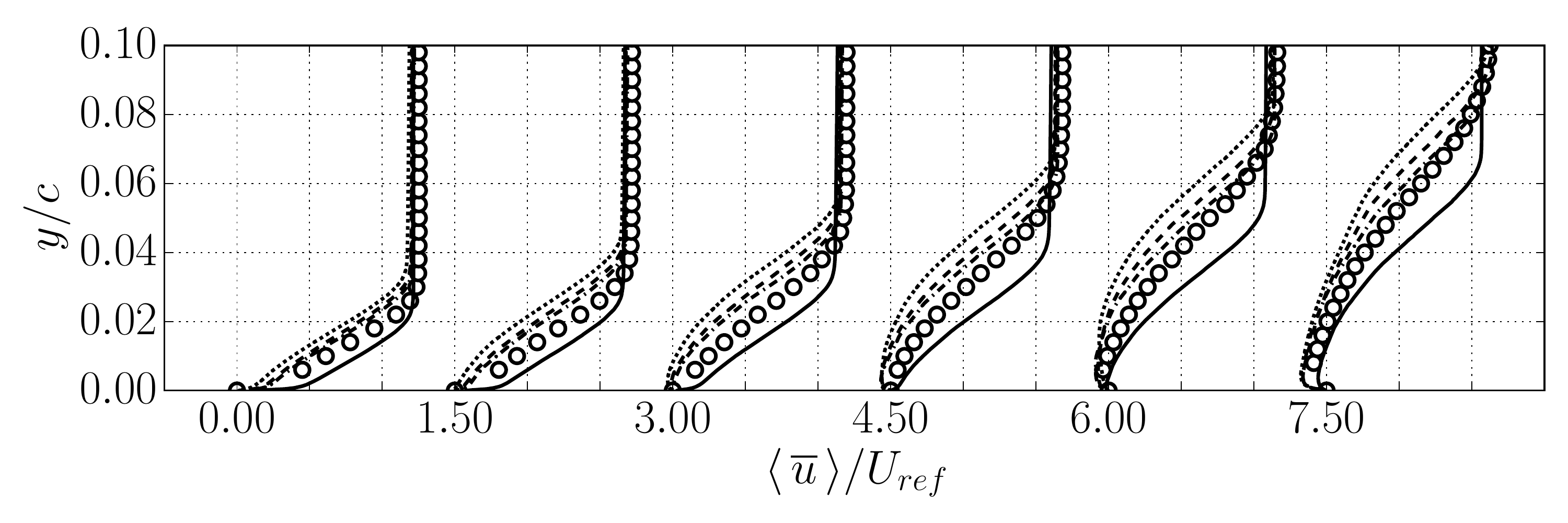}
    \caption{Streamwise velocity profiles in the separated region at locations 
     $x=0.675c$, $x=0.731c$,$x=0.7860$, $x=0.842c$,$x=0.897c$, $x=0.953c$ (from left to right). Symbols are experimental values by \cite{coles1979flying} and lines are from large-eddy simulations for \legendbaseline, (\dotted) case A05(M=0.3). \label{fig:validation_case_u_and_v}}
\end{figure}

The different grid resolution leads to modifications of the friction coefficient $C_f$, which directly translate to different evolution of the boundary layer thicknesses and localization of the separation line. It highlights the high sensitivity of the separation with the grid resolution: slight under-resolution of the
friction may result in a reduced boundary layer thickness and an associated displacement of the separation point further downstream. In this region of maximum lift region, \cite{wadcock1978flying} showed that a small change of angle of attack from $\alpha=12^{\circ}$ to $\alpha=14^{\circ}$ may shift the measured separation point by about 30\% of the airfoil chord. 
The resolution in wall units is described in the figure \ref{fig:Cp_validation}b. None of the grids are adequately resolving the upstream turbulent boundary layer, but B and C grids presents a very good wall resolution in the 30-40\% of the chord length preceding the separation point, while this region is reduced to approximately 10\% with the A grid.
We thus experience a large grid sensitivity of the separation point localization due to the wall friction estimation, as described in the \fig{fig:validation_case_u_and_v}). With the finest grid, the boundary layer thickness is very well predicted (\fig{fig:validation_case_u_and_v}) and the associated horizontal velocity profile $\langle\,\overline{v}\,\rangle$'(\fig{fig:validation_case_u_and_v}a) is overall well captured although slightly underestimated for three stations close to the trailing edge at the stations $x= 0.8418c , 0.8973c ,$ and $ 0.9528c$.  This small error is consistent with the over prediction of the pressure coefficient and leads to an earlier separation. 


\subsection{Extension to compressible subsonic regime ($M_{\infty}=0.3$)}

To release the acoustic CFL stability constraint induced by the explicit time-stepping in \cx~and significantly reduce the cost of the simulations, we set the free-stream Mach number at a higher value ($M_{\infty}=0.3$) compared to the experimental conditions ($M_{\infty}=0.08$), and define the \textit{baseline case}.
This Mach number increase is mild enough to prevent supersonic flow and shock waves, even in the strong acceleration region at the leading edge. Furthermore, it has been shown \cite{ScaloBL_PoF_2015} that higher response from IBC may be obtained with Mach number increase.
We also choose the coarsest grid level A05 to further decrease the computational cost, as most of the flow features are captured using this grid. We thus choose to significantly under-resolve
the near wall turbulence and associated friction in the region [$0.05c-0.5c$], while we promote an appropriate resolution of the outer part of the boundary layer and the shear layer in the separated region.
The higher Mach number choice directly affects the local Reynolds number by modifying the density and results in a thicker boundary layer, associated with a separation point closer to the leading edge.  Although the
grid resolution is marginal,  we obtain a flow field relevant to trailing edge separation. The rather coarse resolution definitely
induces modification in the boundary layer thickness  and flow separation localization, but provides a sufficiently realistic flow field for the purpose of the current study. The airfoil boundary layer interaction with Impedance Boundary Conditions is indeed studied in a regime which does not target the modification of very small scale turbulence developing in the immediate vicinity of the viscous sublayer. 
 


\section{Design of Control Strategy via Impedance Boundary Conditions} \label{sec:control_strategy}

\begin{figure}
\centering
\includegraphics[width=0.85\textwidth]{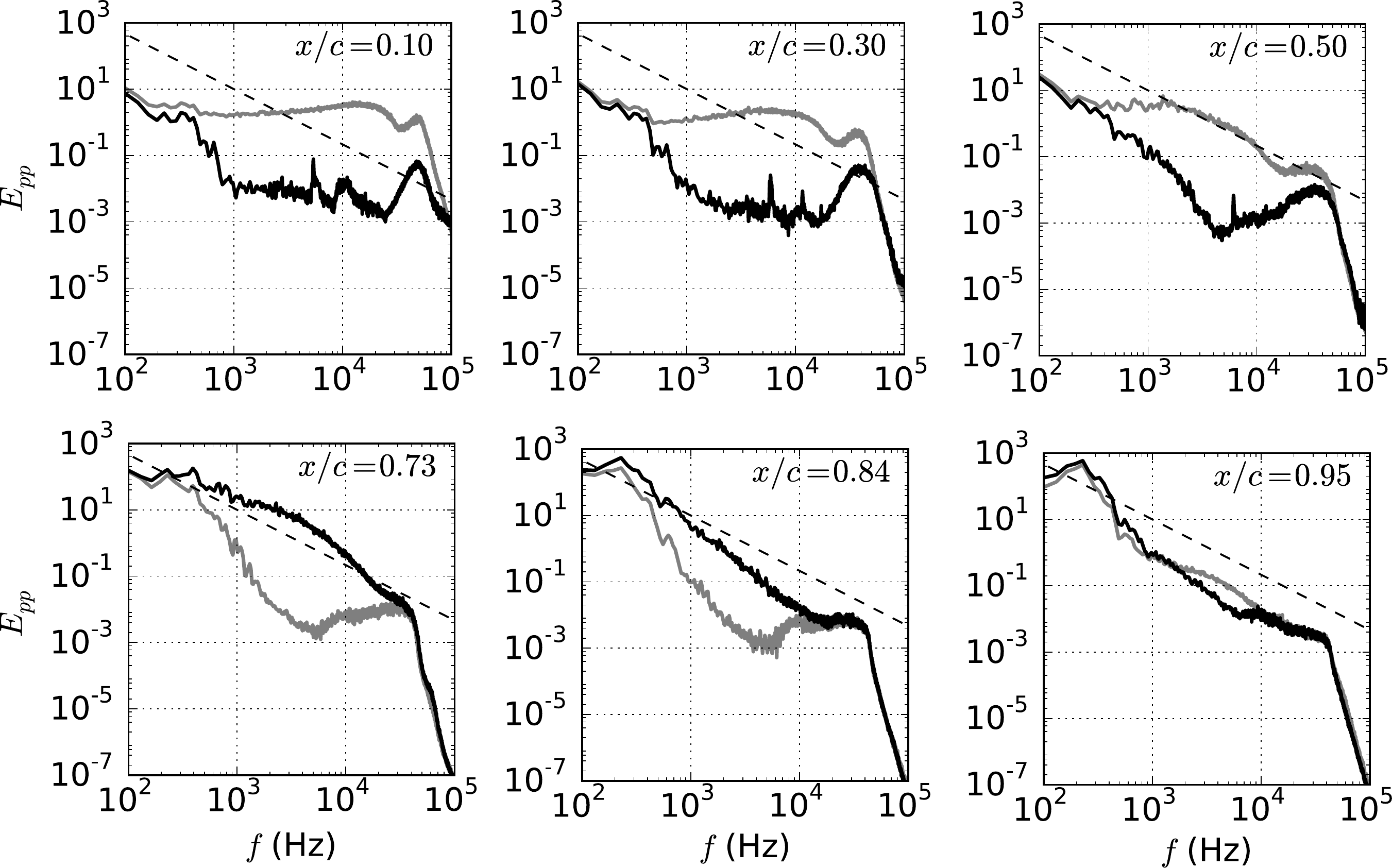}
\caption{Temporal power spectrum of pressure fluctuations at various stations along the suction side of the airfoil in uncontrolled conditions, near the wall (gray) and at approximately one boundary layer thickness away from the wall (black).
}
\label{fig:baseline_pressure_spectra}
\end{figure}

\begin{figure}
\includegraphics[width=\linewidth]{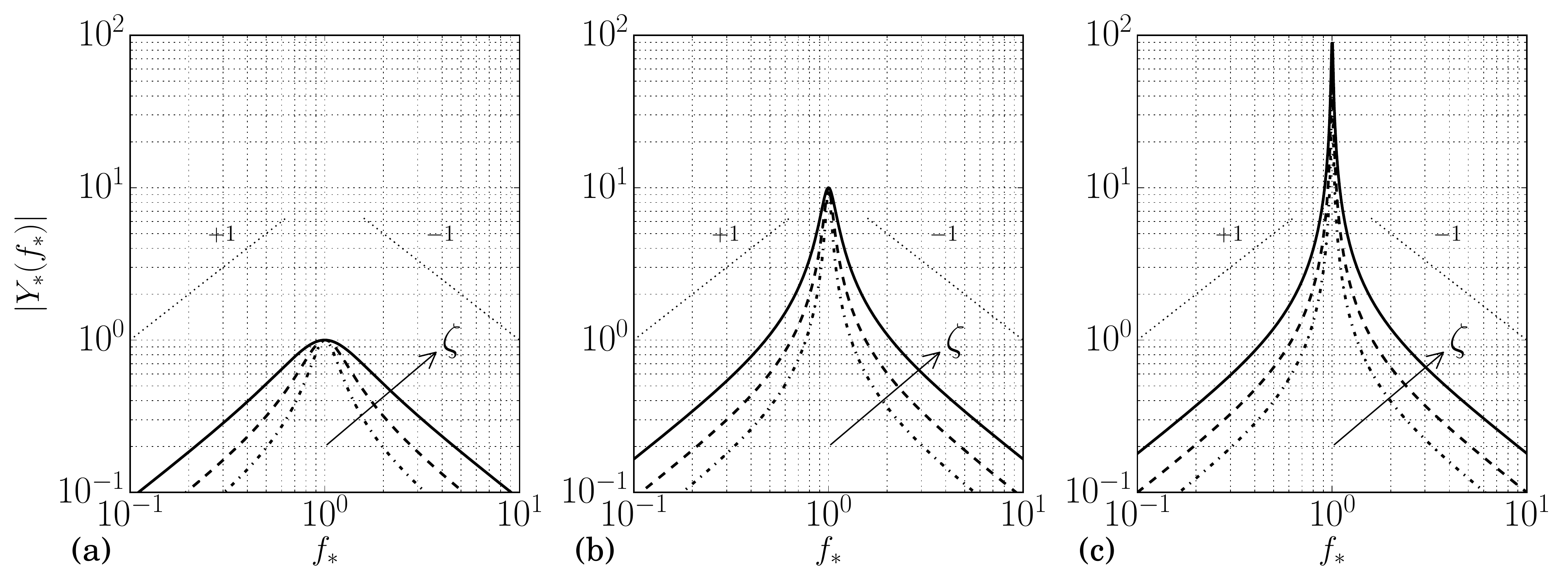}
\caption{Magnitude of admittance, versus dimensionless frequency (inverse of \eqref{eq:single_oscillator_IBC}) for $R=1.00$ (a), $R=0.10$ (b) and $R=0.01$ (c) for $\zeta=0.3,0.5,0.9$ versus frequency normalized by the dimensionless undamped resonant frequency, $f_*=f/f_\textrm{res}$.}
\label{fig:dimensionless_admittance}
\end{figure}

The IBCs used are those of a single-pole Helmholtz oscillator,
\begin{equation} \label{eq:single_oscillator_IBC}
Z(\omega) = R + i\left[\omega\,X_{+1} - \omega^{-1}X_{-1} \right],
\end{equation}
where $R$ is the resistance (dimensionless) and the $X_{+1}$ and $X_{-1}$ are the acoustic mass and stiffness, respectively, and $\omega$ is the angular frequency. The adopted impedance model represents a mass-spring-damper-like second-order system or, alternatively, a frequency-selective porous surface, with 
\begin{equation} \label{eq:tuning_conditions}
f_\textrm{res} = \frac{1}{2\pi} \sqrt{ \frac{X_{-1}}{X_{+1}} }
\end{equation}
 and two other degrees of freedom: the resistance, $R$, inverse of peak maximum attainable permeability; and the damping ratio $\zeta$, regulating the bandwidth of the frequency response.

In preliminary trials several locations of the impedance panel were tested, targeting the boundary layer on the suction side at different stages of development, from 
strong favorable pressure gradient (near the trip), adverse pressure gradient upstream of the separation zone and fully immersed in the separated region, $x>x_{sep}$ (\fig{fig:naca4412}). When a sufficiently strong background mean shear is present, the flow response is manifest in the form of a hydrodynamic Kelvin-Helmholtz (KH) instability triggered by wall-normal acoustic resonance modeled by the imposed IBCs.
The investigation hereafter focuses on analyzing the sensitivity of the flow response to various IBC parameters by fixing the impedance panel to the thus-found sub-optimal location (Fig.~\ref{fig:naca4412}) for which the actuation is located right before the separated region where the generated KH rollers periodically reenergize the otherwise separated boundary layer.
Consistently with \cite{huang1988effect}, resonant frequencies were varied in the range $1/3\,F^+_\textrm{shed} < F^+_\textrm{res} < 3.5\,F^+_\textrm{shed}$ where $F^+_\textrm{shed}$  is the shedding frequency of the uncontrolled flow, identified as the peak of the pressure fluctuation spectrum in the detached region (not shown here) and normalized using $l_\textrm{sep}, U_{\infty}$ to form a reduced frequency commonly adopted by the \emph{active} flow control community. The resistance was varied in the range $0.1--0.5$, bracketing the value $R \sim 0.2$, measured for acoustic liners \cite{TamA_AIAA_1996}. All the cases are synthesized in the table~\ref{table:ibcpanels_2} and have been computed using the A grid described earlier.

\newcommand{\subtablefreq}[2]{
\begin{tabular}{l}
\vspace*{0.1cm}
$\fpres=#2$ \\
$\sttetres=#1$ \\
\end{tabular}
}

\newcommand{\subtableres}[1]{
\begin{tabular}{l}
\vspace*{0.1cm}
$R=0.1$ \\
$R=0.2$ \\
$R=0.5$ \\
\end{tabular}
}

\renewcommand{\arraystretch}{1.7}
\begin{table}
\centering
\begin{tabular}{c|c|c|c|c}
 &\subtablefreq{0.0041}{0.14}  &   \subtablefreq{0.0082}{0.28} &  \subtablefreq{0.019}{0.67} &  \subtablefreq{0.038}{1.34} \\
\cline{1-5}
Cases
  &\subtableres{F1}  &   \subtableres{F2} &  \subtableres{F3} &  \subtableres{F4}\\
\cline{1-5}
 \end{tabular}
 \caption{Panel settings for the twelve controlled cases: resonant frequency $\ffres$, resistance $R$. For all cases, the damping ratio $\zeta=0.5$ and panel location is  $x_{p0}$--$x_{p1}$.
 } \vspace*{0.2cm}
\label{table:ibcpanels_2}
\end{table}



\newpage

\section{Results: Performance of Passive Acoustic Control} \label{sec:results}

The flow modifications observed in this particular setup are similar to those obtained in the channel flow configuration discussed in \cite{ScaloBL_PoF_2015}, although we address here a different range of resonator frequencies and associated length scales $U_{\infty}/\fres$ greater than the boundary layer thickness $\delta$.\\
Cases for which the IBC panel induce a strong response are characterized by the generation of a downstream travelling wave of transpiration velocity (\ie normal to the panel) in the vicinity of the panel. 

\begin{figure}
\centering
\includegraphics[width=\textwidth]{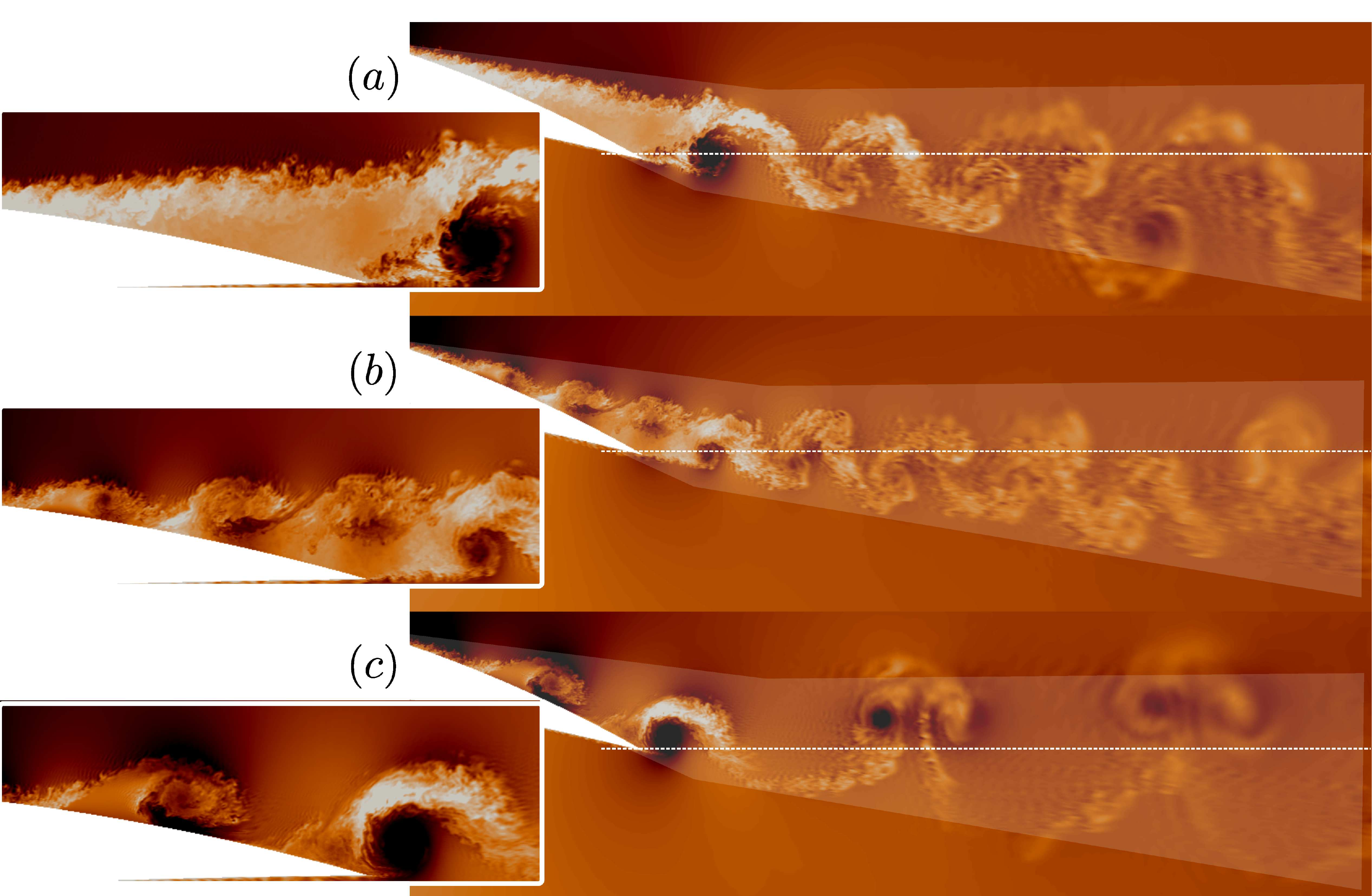}
\caption{
Vizualisation of temperature, acting as a passive scalar in the wake of the \nacafft~airfoil for (a) baseline, (b) \fthreerone~and (c) \ftworone. Transparency area is used as an indicator for the baseline wake main features. \label{fig:visu}}
\end{figure} 

This observation is consistent with the hydro-acoustic instability previously identified in \cite{ScaloBL_PoF_2015} and is the dominant mechanism for all the cases which presents a strong response. 
i) the resonant frequency is excited through the broadband turbulent structures.
ii) the resulting oscillating transpiration velocity provides both an inflectional velocity profile and an initial perturbation to trigger Kelvin Helmholtz instability,
which iii) in turns provides back a fluctuating pressure level associated with the flow structure dominated by successive rollers advected downstream iv) which resonates through the IBC panel and sustain the excitation of the shear layer.

The intensity of the response is measured using the $RMS$ of the transpiration velocity $v_n^{\prime}$ which may become large in some  cases, and comparable to the free stream velocity, with $v_n^{\prime}\approx 0.25 U_{\infty}$.
To characterize the amplitude of the response we adapt the definition of the momentum coefficient(see \eg~\cite{glezer2005aspects}): 
\begin{equation}
C_\mu  = \frac{2\rho_j\;\int_\textrm{panel} {v_n^\prime}^2 dx}{\rho_\infty\,c\, U_\infty^2}
\end{equation}
Within this framework, $C_\mu$ is known \emph{a posteriori} and gives a measure of the response amplitude rather than emphasizing the cost of an \emph{active} control framework. The transpiration velocity is measured and integrated over the impedance panel at the first off-wall cell. Values of $C_\mu$ are reported in the table~\ref{tbl:ibc_settings}. Not surprisingly, higher permeability leads to larger response. For a given permeability the response amplitude is maximal for the \ftwo~frequency. This suggests a maximal response as we approach the shedding frequency of the shear layer, which is consistent with the unstable nature of the shedding mode.
Thus, the resonant frequency sets the time scale of the shear layer, while the amplitude of the transpiration velocity for a given permeability is directly dependent of the surrounding instabilities natural frequencies. The combination of the time scale $f_\textrm{res}$ and the free-stream velocity $U_\infty$ provides the length scale $l_{KH,0}$ of the Kelvin Helmholtz rollers developing in the shear layer. Visualization of the instantaneous flow field in the figure~\ref{fig:visu} confirm that $l_{KH,0}\sim U_\infty/f_\textrm{res}$.

\begin{figure}
\centering
\includegraphics[height=0.32\textwidth]{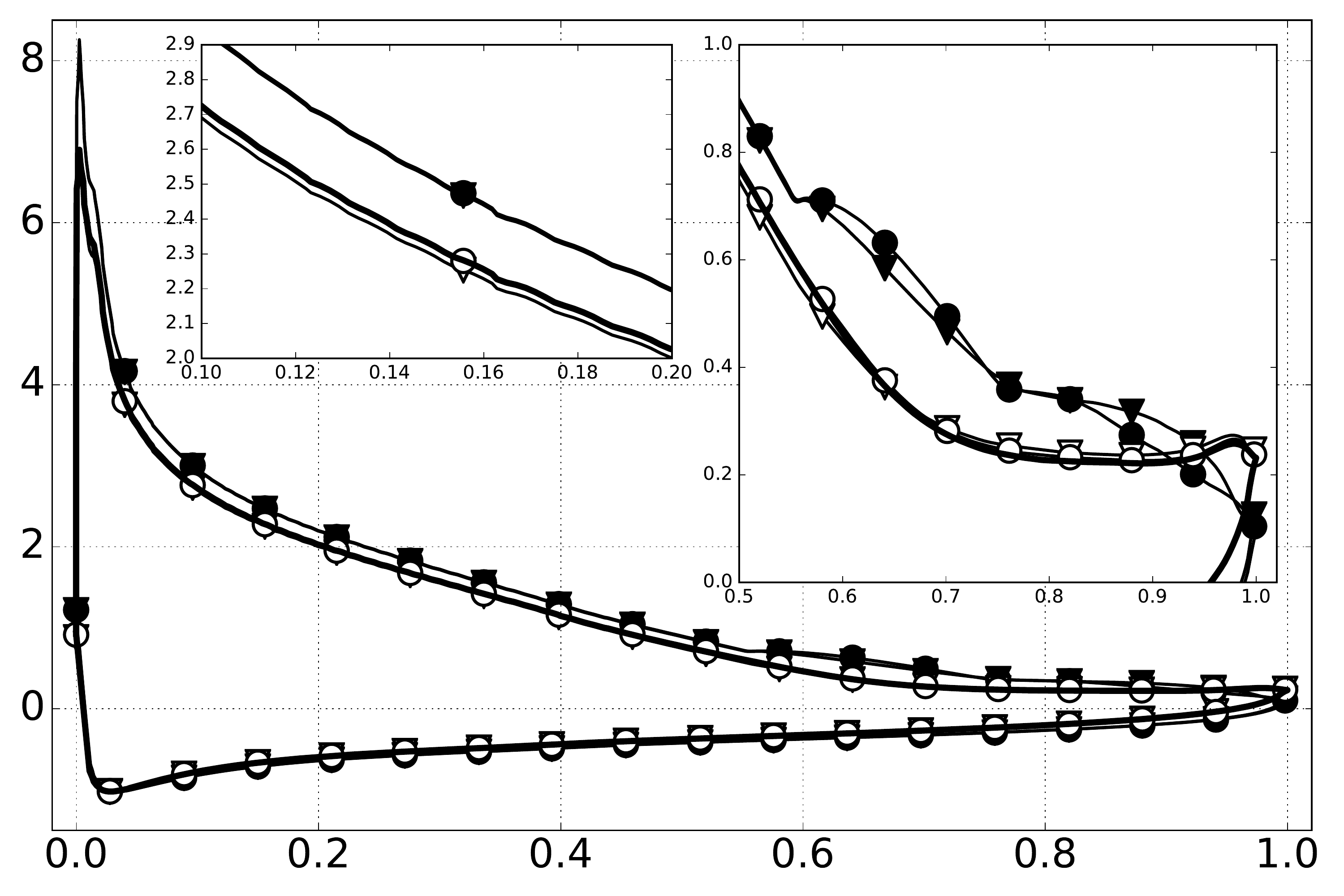}
\includegraphics[height=0.32\textwidth]{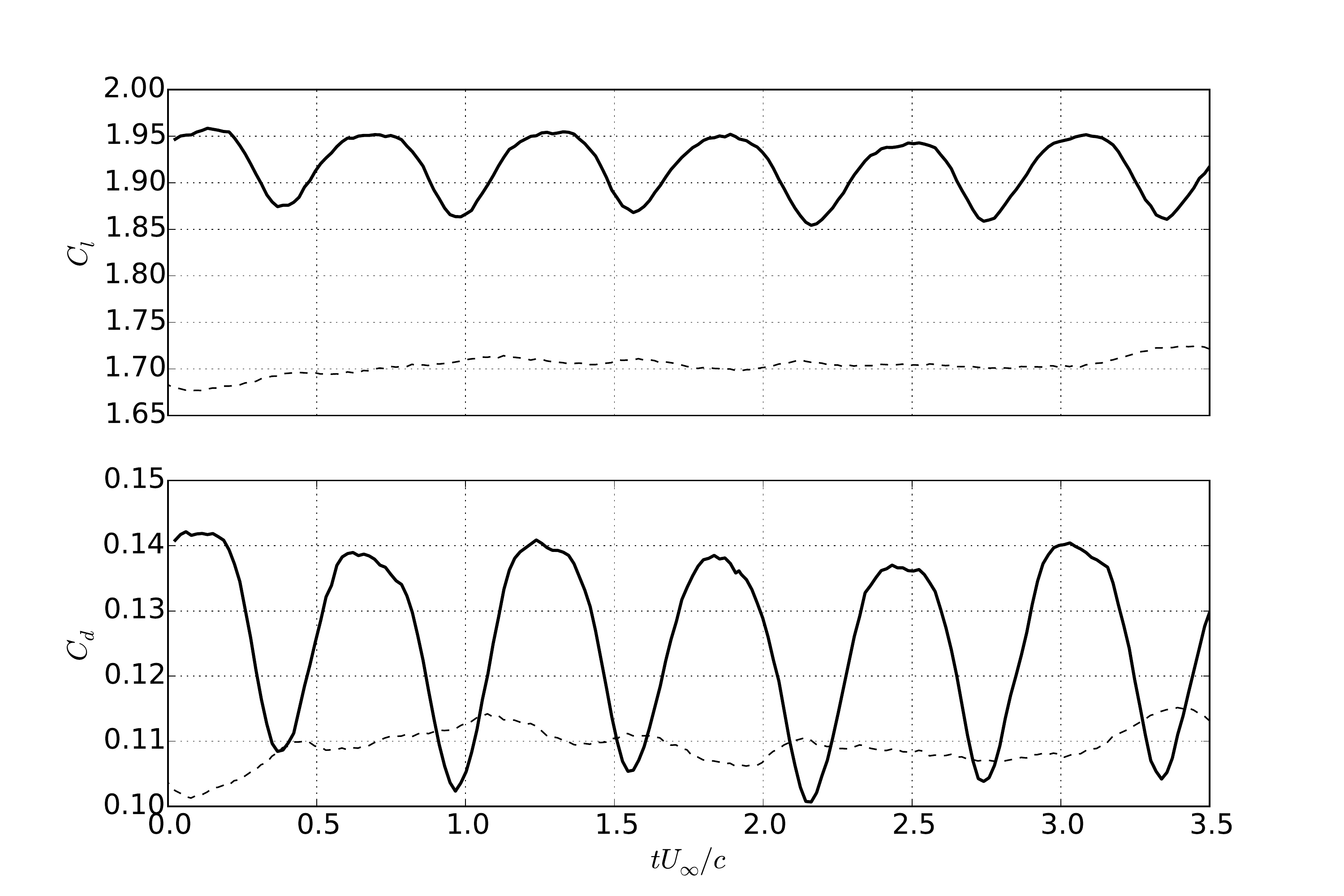}
\caption{ (a)Evolution of the pressure coefficient $-C_p$ near the trailing edge for $R=0.2$ \legendone and (b)
Unsteady pressure contribution to the lift and drag coefficients for cases \ftwortwo~(\solid) and \ftworfive(\dashed).
\label{fig:cpcl}
}
\end{figure}

To quantify the performance of the control, we analyze  the integrated aerodynamic coefficients $C_l$ and $C_d$, summarized in the table~\ref{tbl:ibc_settings}. We observe a net drag increase even for low permeability values which is a direct consequence of the enhanced turbulent activities owing to the impedance panel. Regarding lift, a net increase is observed for several cases, 
therefore showing successful control of the separated flow.
The local pressure coefficient $C_p$ shown in the figure \ref{fig:cpcl}, reveals the main contributions to this lift increase.
First, the large scale structures contribute to vortex induced lift: low pressure vortex cores are convected over the suction side of the airfoil.  Second, a close view of the  upstream region of the boundary layer indicates a global lift increase, attributed to the shear layer reattachment between each KH rollers and the complete removal of the recirculation zone in the mean flow streamlines.
The first contribution is the main unsteady component of the total lift, as shown in the figure \ref{fig:cpcl}b. 
The four cases showing significant increase suggests the following mechanism that drives a successful control using acoustic resonance:
i) successful control occurs if the generated structure sizes $l_{KH}$  at the trailing edge is larger or of the order of the baseline separation height: in this situation, momentum mixing is strongly enhanced and high momentum fluid is drained towards the wall as illustrated in the figure \ref{fig:visu}.
ii) two ingredients are required to trigger the initial development of the forced $KH$ instability: the mean shear, provided by the upstream boundary layer and a threshold amplitude of the excitation through the panel response. The current study shows that the threshold amplitude may be reached with reasonable permeability levels, provided the resonant frequency of the Helmholtz resonator is close to the shear layer shedding frequency.

Note that for successfully controlled cases, the generated large scale structures directly interact with the wall in the otherwise separated region. At the trailing edge a counter rotating vortex is formed from the pressure side as the result of this close interaction. It results in a very different and strongly asymmetric wake. This variation of topology in the wake may be leveraged as an indicator of successful control as it is the footprint of the trailing edge interaction.

\renewcommand{\arraystretch}{1.1} 
\begin{table}
\centering
\begin{tabularx}{\textwidth}{m{2.5cm}m{2.5cm}m{2.5cm}m{2.5cm}m{2.5cm}}
&\fone & \ftwo & \fthree & \ffour \\
$C_\mu(\%)$ &  \\
\hline
$R=0.1$ & 2.49e-01 & 4.93e-01 & 1.62e-01 & 1.30e-02 \\
$R=0.2$ & 6.01e-02 & 1.49e-01 & 8.78e-03 & 4.08e-03 \\
$R=0.5$ & 5.48e-03 & 3.55e-03 & 3.53e-03 & 3.33e-03 \\
$C_l$ &  \\
\hline
$R=0.1$ &  1.77 (+10\%) & 1.76 (+9\%) & 1.65 (+3\%) & 1.59 (-1\%) \\
$R=0.2$ &  1.77 (+10\%) &  1.82 (+13\%) & 1.62 (+1\%) & 1.61 ($\sim$0\%) \\
$R=0.5$ & 1.63 (+1\%) & 1.63 (+1\%) &  1.61 ($\sim$0\%) & 1.61 ($\sim$0\%) \\
$C_d$ &  \\
\hline
$R=0.1$ &  0.0723 (+64\%)  &  0.0836 (+90\%) & 0.0509 (+16\%) &  0.0529 (+20\%) \\
$R=0.2$ &  0.0653 (+48\%)  &  0.0657 (+49\%) &  0.0523 (+19\%)  & 0.0484 (+10\%) \\
$R=0.5$ &  0.0535 (+22\%)  &  0.0497 (+13\%) & 0.0479 (+9\%) & 0.0479 (+9\%) \\
$C_l/C_d$ &  \\
\hline
$R=0.1$ &  25 (-34\%) &  21 (-43\%) & 32  (-12\%)& 30 (-19\%) \\
$R=0.2$ &  27 (-27\%)   &   28 (-25\%)&  31(-16\%) &  33(-10\%) \\
$R=0.5$ &  31 (-18\%) &  33  (-12\%)& 34  (-9\%)&  34 (-9\%) \\
\end{tabularx}
\caption{Momentum coefficient $C_\mu$ characterizing the response of the impedance panel. Lift $C_l$ and drag $C_d$ coefficients for various frequency and values of resistance. Drag coefficient $C_d$ is estimated from wake survey method and lift coefficient $C_l$ is estimated from $C_p$ integration. \label{tbl:ibc_settings} }
\end{table}


\section{Conclusions}

Passive flow control on a \nacafft~airfoil was investigated by means of acoustic resonance of modeled Helmholtz resonators. We manage to achieve successful hydrodyanmic control of the trailing edge flow separation, when appropriate impedance panels are located just upstream the separated region.
Although the control set up is fully passive, it operates at the natural resonator frequency, and may be tuned differently to target different conditions of operation. We show that a significant (+13\%) increase in lift is obtained if the resonant frequency is appropriately chosen regarding the time scales of the detached shear layer: frequencies below the shedding frequency $F^+_\textrm{res}<F^+_\textrm{shed}$ used in combination  with relatively high permeability ($R=0.2$) generate large coherent Kelvin Helmholtz rollers, sufficient to drain momentum from the farfield and periodically reattach the flow field. This
passive flow control strategy is very attractive as it may address different ranges of operation, hence offering passive set up advantages of reduced maintenance and higher reliability, while providing additional flexibility in comparison with traditionnal passive set up.

\bibliography{./references1,./references2,./references_scalogroup,NACA4412}
\bibliographystyle{aiaa}

\end{document}